\documentclass[prl,twocolumn,showpacs,amsmath,amssymb,superscriptaddress,floatfix]{revtex4}

\usepackage[english]{babel}
\usepackage{amsfonts}
\usepackage{graphicx}

\begin{document}

\title{Bipolaron in the $t-J$ model coupled to longitudinal and transverse quantum lattice vibrations}

\author{L. Vidmar}
\affiliation{J. Stefan Institute, 1000 Ljubljana, Slovenia}

\author{J. \surname{Bon\v ca}}
\affiliation{Faculty of Mathematics and Physics, University of Ljubljana, 1000
Ljubljana, Slovenia}
\affiliation{J. Stefan Institute, 1000 Ljubljana, Slovenia}

\author{S. \surname{Maekawa}}
\affiliation{Institute for Materials Research, Tohoku University,
Sendai 980-8577, Japan}
\affiliation{CREST, Japan Science and
Technology Agency (JST), Tokyo 102-0075, Japan}

\author{T. \surname{Tohyama}}
\affiliation{Yukawa Institute for Theoretical Physics, Kyoto
University, Kyoto 606-8502, Japan}


\date{\today}
\begin{abstract}
We explore the influence of two different polarizations of quantum oxygen vibrations   on the spacial symmetry of the bound
magnetic bipolaron in the context of the $t-J$ model by using
exact diagonalization within a limited functional space. Linear as well as  quadratic electron phonon coupling to transverse polarization    stabilize $d-$wave symmetry. The existence of a magnetic background is essential  for the formation of a $d-$wave bipolaron state. With increasing
linear electron phonon coupling to longitudinal  polarization  the symmetry of a $d$-wave
bipolaron state changes to a $p$-wave. Bipolaron  develops  a large anisotropic effective mass.
\end{abstract}

 \maketitle

Soon after the discovery of high-$T_c$ superconductivity the
quest for the pairing mechanism focused on magnetic fluctuations
due to a   broadly   accepted conjecture 
that  phonon mechanism alone is not strong
enough to produce high transition temperatures as observed in
high-$T_c$ compounds. 
Recently, a growing evidence is emerging 
in favor of the significance  of lattice degrees
of freedom in high-$T_c$ compounds \cite{lanzara1,alexandrov,newns}.
The interplay between
strong correlations and lattice degrees of freedom \cite{gunnarsson} seems to be
responsible for many unusual properties of cuprates in the low
doping regime, such as kinks \cite{lanzara1}, stripes \cite{reznik},  and 
waterfall \cite{ronning,bonca3} structures. 

A long-standning objection against phonon-based mechanism for high-$T_c$ superconductivity is based on a widely  accepted notion that  coupling to phonon degrees of freedom is predominantly consistent with $s-$wave pairing, not characteristic for cuprates.  Despite recent discovery that weak EP coupling to acoustic phonons in the presence of   large on-site Coulomb interaction  leads to $d-$wave pairing \cite{alexandrov1}, the role of short-wavelength oxygen oscillations on the symmetry of the paired state remains a challenging open problem.

Investigation of correlated models coupled to phonons were based on  exact diagonalization (ED) calculations on small lattice systems \cite{riera_stripe,sakai,peter,fehske}, slave-boson approaches \cite{ramsak2,kyung,ishihara}, dynamical
mean-field calculations \cite{sangiovanni,macridin,cappelluti}, coherent states Lanczos method \cite{mishchenkoNONLOC},  and quantum Monte Carlo
methods (QMC) \cite{nagaosa,fehske}. In Ref.~\cite{riera_stripe} authors present a detailed study of the influence of EP coupling to a single phonon mode  on formation of inhomogeneous charge structures in the $t-J$ model. They  show that half-breathing mode stabilizes a stripe phase. In contrast, using  slave-boson approach authors of  Ref.~\cite{ishihara} suggest, that half-breathing mode  enhances $d-$ wave pairing. { They furthermore  underline the importance of off-diagonal EP coupling modulating the  hopping and  the spin-exchange terms. In contrast, authors of Ref.~\cite{gunnarsson1} find that diagonal EP terms exceed off-diagonal ones by nearly two orders of magnitude. 
}

ED calculations of the $t-J$
model  show that the $d-$wave symmetry
of a bipolaron \cite{cherny1,eder,riera2,barentzen,footnote}  is not robust against addition
of longer range hopping terms \cite{leung2} while  recent QMC calculations of the Hubbard model yield  $T_c$  far below those of cuprates \cite{aimi}. There seems to be a need to uncover additional  mechanism that would help stabilize the $d-$wave symmetry of a bound bipolaron state. In this Letter we show that EP coupling to a transverse  polarization (TP) of oxygen vibration  provides an important  mechanism  that stabilizes  the $d-$wave symmetry of a  bound hole pair  with a small effective mass. 

\begin{figure}[htb]
\includegraphics[width=8cm,clip]{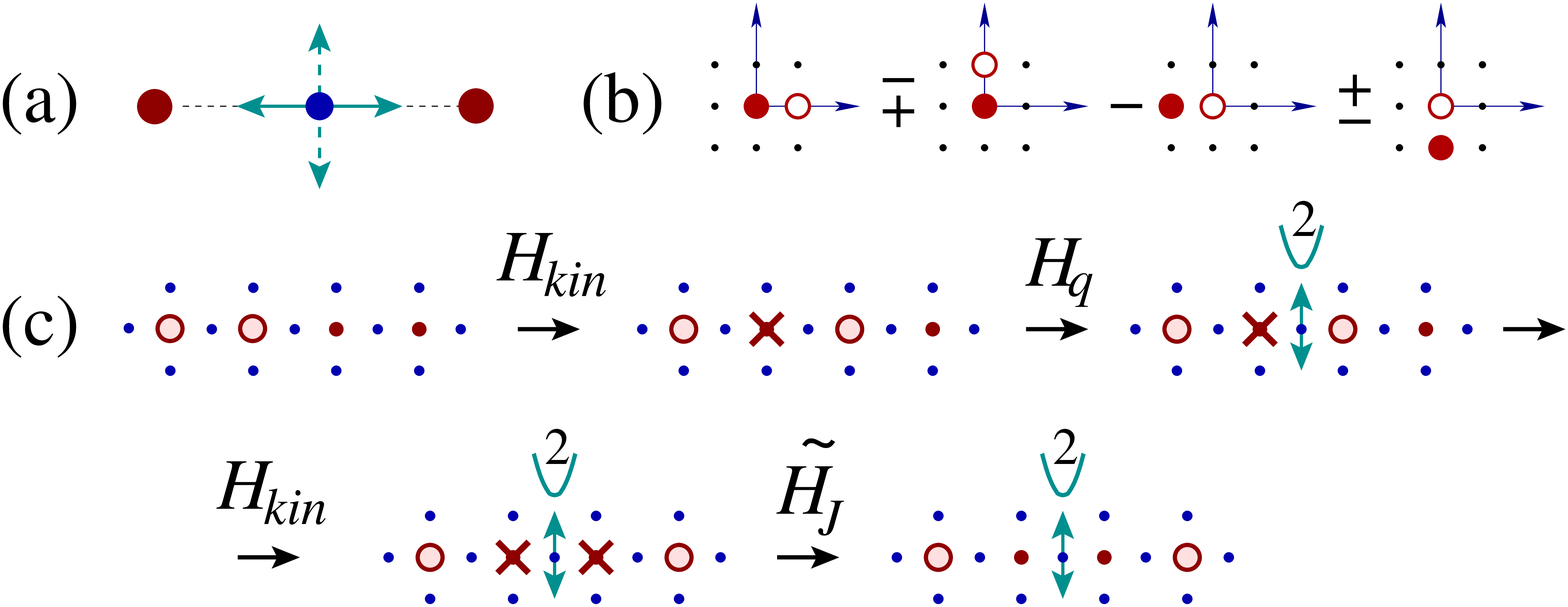}
\caption{(Color online) %
(a) Schematic representation of  LP and TP vibrations of O atom
(middle) with respect
to Cu-O-Cu bond, (b) schematic representation of a $d_{x^2-y^2}-$wave ($s-$wave) top (bottom)
signs two-hole starting wavefunction. Fermion sign convention
places the first hole depicted with the full circle to the left-most, if
the pair is vertical, then bottom-most position. Only Cu sites are
presented with small dots in (b); (c) schematic representation of succeeding applications
of different off-diagonal parts of Hamiltonian in Eq.~\ref{ham}, for $n_h=4$,  starting
from a single hole-pair in the N\' eel state with zero phonon quanta.
Dots represent Cu and O atoms, holes are denoted by open circles,
crosses represent spin-flips (overturned spins with respect to the original
N\' eel configuration of spins, localized on Cu sites), vertical arrows
indicate and point to the  numbers of  excited TP phonon quanta.
}\label{fig1}
\end{figure}

We   solve a system of two holes in the $t-J$ model defined on an
infinite two-dimensional lattice by extending    the  method for a single hole based  on exact diagonalization within a limited functional space \cite{bonca3,bonca2}.
%
%
We introduce diagonal EP  coupling to either TP of 
oxygen (O) vibration  relevant for the description of buckling modes or longitudinal polarization (LP) of O vibration  relevant for description of  bond-streching modes. 
We investigate the following Hamiltonian
\begin{eqnarray}
H&=& -t\sum_{\langle {\bf i,j}\rangle,s}(\tilde c^\dagger_{{\bf i},s} \tilde c_{{\bf j},s} +\mathrm{H.c.}) +
J\sum_{\langle {\bf i,j}\rangle }( {\bf S}_{\bf i} {\bf S}_{\bf j} - \frac{1}{4}n_{\bf i} n_{\bf j} ) \nonumber \\
  &+& g \sum_{\bf i,\boldsymbol{\delta}}(n_{\bf i}^h - n_{\bf i+\boldsymbol{\delta}}^h)(a_{{\bf i} +\boldsymbol{\delta}/2}^\dagger + a_{{\bf i} + \boldsymbol{\delta}/2})\nonumber \\ 
  &+& q_\beta \sum_{\bf i,\boldsymbol{\delta}}(n_{\bf i}^h + n_{\bf i+\boldsymbol{\delta}}^h)(a_{{\bf i}+\boldsymbol{\delta}/2}^\dagger + a_{{\bf i} + \boldsymbol{\delta}/2})^\beta \nonumber \\
&+&\omega_0\sum_{{\bf i} + \boldsymbol{\delta}}a_{{\bf i}+\boldsymbol{\delta}/2}^\dagger  a_{{\bf i}+\boldsymbol{\delta}/2},\label{ham}
\end{eqnarray}
where $\tilde c_{{\bf i},s} = c_{{\bf i},s}(1 - n_{{\bf i},-s})$ is a projected  fermion operator, $t$ represents nearest neighbor overlap integral, the sum $\langle \bf i,j \rangle$  runs over pairs of nearest neighbors, $a_{\bf i}$  are phonon annihilation operators and $n_{\bf i} = \sum_s n_{{\bf i},s}$. The third term represents linear EP  coupling to LP of  O vibration with respect to Cu-O-Cu bond, see also Fig.~\ref{fig1}(a). { Fourth term is chosen either linear $(\beta = 1)$ or quadratic $(\beta = 2)$ in O displacement, describing TP of  O vibration, Fig.~\ref{fig1}(a).  $\beta =2$ is chosen to describe the  CuO plane  with no pre-buckling of O positions.} Sums over $\boldsymbol{\delta}$  in the latter two terms run over two orthogonal nearest neighbor Cu positions.  Lattice vibrations on  O sites are independent - we do not predispose any particular phonon mode 
with the exception of limiting our calculation to  either TP or LP of O oscillation. In treating quantum  phonons we follow well established approach of Ref.~\cite{bonca1}.

The construction of the functional  space starts from a N\' eel state with two holes located on neighboring Cu sites and with zero phonon quanta. Such a state represents a parent state of a translationally invariant state with a given momentum $k$. In the case of a high symmetry point ${\bf k} = (0, 0)$ the parent state can be chosen to have $d-$, $s-$, or $p-$ wave symmetry as  for the case of $d-$ and $s-$  shown   in Fig.~\ref{fig1}(b). The starting state is  written as 
%
$%
\vert \phi^{(0)}{\rangle}_a = \sum_{\boldsymbol{\gamma}}(-1)^{M_a(\boldsymbol{\gamma})}
c_0c_{\boldsymbol{\gamma}}\vert {\rm Neel };0\rangle,
$
%
where sum runs over four nearest neighbors in the case of $d-$ and $s-$ wave symmetry and over two in the case of $p_{x(y)}$-wave while $M_a (\boldsymbol{\gamma})$, $a\in \{ d,s,p \}$ sets the appropriate sign.

We generate new parent states by applying the generator of states 
$
\left \{ \vert \phi_l^{(n_h)}{\rangle}_a\right \} = \left ( H_{\rm kin} + \tilde H_J + 
H_{\rm ph}\right )^{n_h} \vert \phi^{(0)}{\rangle}_a;~n_h=1,\dots,N_h
$ where
$H_{\rm kin}$  represent the first term in Eq.~\ref{ham}, $\tilde H_J$ denotes a part  of the  second
term in Eq.~\ref{ham}  which is only applied  to erase spin flips that
were generated through succeeding application of $H_{\rm kin}$, as for
a particular case depicted in Fig.~\ref{fig1}(c). $H_{\rm ph}$  represents either
third or fourth term in Eq.~\ref{ham}.  
This procedure
generates exponentially growing basis of states, consisting
of different shapes of strings in the vicinity of the hole
with maximum lengths given by $N_h$ as well as phonon quanta
created along paths of both holes. { Identical basis functions, generated by different processes, are chosen only once.}  
We have used
$N_h = 8$ that lead to $N_{\rm st} = 13 \times  10^6$ states. Full Hamiltonian
in Eq.~\ref{ham} is diagonalized within this limited functional
space taking explicitly into account translational symmetry.

In Fig.~\ref{fig2} we present the energy difference $ E_p - E_d $ between the lowest $p-$ and the $d-$wave state for two different values of $J/t$ as a function
of $q_\beta/t$ and  $g/t$ for the case of TP and  LP respectively. At $J/t = 0.1$ the two-holes are unbound at $q_{1,2} = g = 0$ \cite{cherny1,eder} and degenerate $p-$wave ground state is found, $ E_p - E_d <0$, see Figs.~\ref{fig2}(a) and (b). 
Increasing $q_\beta/t$ and $g/t$ leads to rather surprisingly distinct results. In both cases increasing EP coupling leads to a formation of a bipolaron, as also evident from Figs.~\ref{fig3}(e) and (f) and the discussion later in the text. 
While coupling to TP leads to a formation of a bound state with the $d-$wave symmetry, coupling to  LP  in contrast favors a bound state with the $p-$wave symmetry. This effect is even more pronounced at larger value of $J/t = 0.4$ where at $q_{1,2} = g = 0$ a bound magnetic bipolaron is
already formed \cite{cherny1,eder,riera2} with a $d-$wave symmetry. { By increasing $q_2/t$, $d-$wave symmetry is stabilized.  Increasing linear EP coupling $q_1/t$ leads to an initial increase of   $E_p-E_d$  followed by a decrease, $E_p-E_d\to 0$  around $q_1/t\gtrsim 0.75$ due to a crossover to a strong EP coupling regime, Fig.~\ref{fig2}(a).
In contrast,  linear EP coupling to LP drives even a bound $d-$wave bipolaron state at $J/t = 0.4$ and $g = 0$ to a bound state with a $p-$wave symmetry at $g/t \sim 0.61$, see Fig.~\ref{fig2}(b). 

Effective bipolaron mass  $m_{\alpha \alpha}=t\left [\partial^2E({\bf k})/\partial {\bf k} \partial {\bf k}\right ]^{-1}_{\alpha \alpha}$, computed in its  eigen-directions,  presented in Figs.~\ref{fig2}(c) and (d),   is isotropic in the case of $d-$wave symmetry and anisotropic, with the anisotropy ratio $m_{yy}/m_{xx}\sim 3-10$  in the case of $p-$wave state. At $J/t=0.4$   $m_{\alpha \alpha}$ furthermore shows only a weak increase with $q_2/t$, see Fig.~\ref{fig2}(c).  Even more surprising is the decrease of the effective mass  at $J/t=0.1$ in the regime of a bound $d-$wave bipolaron, {\it i.e.} for $q_2/t \gtrsim 0.5$. Note, that the nonanalytic behavior of $m_{xx}$ is a consequence of the  symmetry change from $p-$ to $d-$ state at $q_2/t\sim 0.5$ as also seen from Fig.~\ref{fig2}(a). 

Focusing on  linear EP coupling to  LP,  $m_{xx}$ at $J/t=0.4$ starts a rapid increase signaling the approach to  strong EP coupling regime just below the transition to the $p-$wave state, around $g/t\sim 0.57$, see Fig.~\ref{fig2}(d).  As the system enters $p-$wave state 
the mass again becomes anisotropic.

\begin{figure}[htb]
\includegraphics[width=7cm,clip]{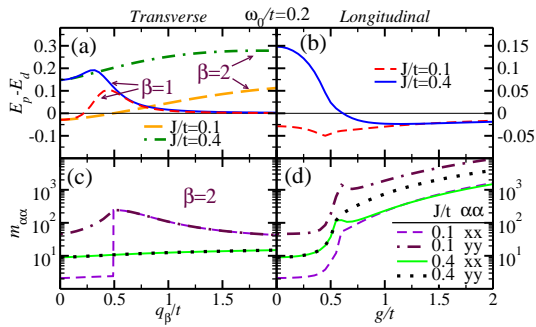}
\caption{(Color online) 
$E_p - E_d$ at $\omega_0/t = 0.2$,  and two
different strengths of $J/t$ vs. $q_\beta/t$ in the case of EP coupling
to TP (a)  and vs. $g/t$ in the case of linear EP coupling
to LP (b). 
Effective masses $m_{\alpha\alpha}$ vs. $q_2/t$ in  (c) and vs. $g/t$ in (d).  The ground state wave-vector is ${\bf k}=0$ except in the regime  $g/t\gtrsim 0.57$
where ${\bf k}=(\pi,0)$.  
}\label{fig2}
\end{figure}

In Fig.~\ref{fig3} we present the  probability of finding a hole-pair at a distance of r: 
$
P(r) = \langle \sum_{\langle \bf i\not = j \rangle } n^h_{\bf i}n^h_{\bf j}
\delta\left [ \vert {\bf i-j}\vert -r\right ]  \rangle
/ \langle \sum_{\langle \bf i\not = j \rangle } n^h_{\bf i}n^h_{\bf j} \rangle,
$ 
and average hole distance
$
\langle d\rangle = \sum_{r} r P(r).
$ 
We first focus on the effect of 
EP coupling to TP, see Figs.~\ref{fig3}(a,c,e). At
$J/t = 0.1$, bipolaron is unbound in the regime { ($q_2/t \lesssim 0.5$, and  $q_1/t \lesssim 0.22$)}, 
nevertheless, $\langle  d \rangle $ remains finite due to a limited Hilbert space
where the maximal inter-hole distance is given by  $l_{\rm max} = N_h +1 = 9$.
Increasing $N_h$ would lead to further increase of  $\langle  d \rangle $ in this
regime as well as to further spread of $P(r)$ towards larger
$r$, see Fig.~\ref{fig3}(c) for $J/t = 0.1$ and $q_{1,2} = 0$. In this range of parameters we observe no exponential decay of $P(r)$, see Fig.~\ref{fig3}(e). 
In contrast, in the regime of a bound bipolaron, i.e.
for $J/t = 0.1$ and { ($q_2/t \gtrsim 0.5$ and  $q_1/t \gtrsim 0.22$)} as well as at $J/t = 0.4$, 
$\langle  d \rangle $ and $P(r)$ 
do not change much with further increasing $N_h$ and exponential decay is clearly observed in  Fig.~\ref{fig3}(e). 
Our tests performed on smaller systems ($N_h = 4$ and 6) reaffirm
that results in the regime of a bound bipolaron have indeed
converged close to a thermodynamic limit. Good agreement
of  $P(r)$ at $J/t = 0.4$ and $q_{1,2} = g = 0$ 
is  found with ED calculation on 32-sites cluster, Ref.~\cite{cherny1,riera2}. Structure of a bound bipolaron, revealed
by $P(r)$ at $J/t = 0.4$ and $q_{1,2} = 0$, is remarkably similar to
that computed at $J/t = 0.1$ and $q_2/t = 1.0$ and $\beta = 2$, Fig.~\ref{fig3}(c).
{ Both, quadratic and linear} EP coupling to TP
lead to a formation of a bound bipolaron with the $d-$wave symmetry even in the
case of small exchange interaction $J/t = 0.1$ where magnetic
mechanism is not strong enough to form a bound magnetic
bipolaron. To investigate whether coupling to TP 
alone can lead to $d-$wave state  in the absence   of
a magnetic background, we have solved a problem with two
spinless particles quadratically coupled  to TP  or linearly to  LP using
 topology of a Cu-O plane. 
By increasing $q	_2/t$ or $g/t$ we obtain in both cases a  bipolaron with a $p-$wave symmetry, Figs.~\ref{fig3}(a)
and (b). 
We thus emphasize an important conclusion: EP coupling to TP stabilizes
$d-$wave symmetry  of  a hole-pair, however, the existence of a
magnetic background as found in the $t - J$ model
seems to be essential precondition for the formation of a
$d-$wave  state. 

\begin{figure}[htb]
\includegraphics[width=7cm,clip]{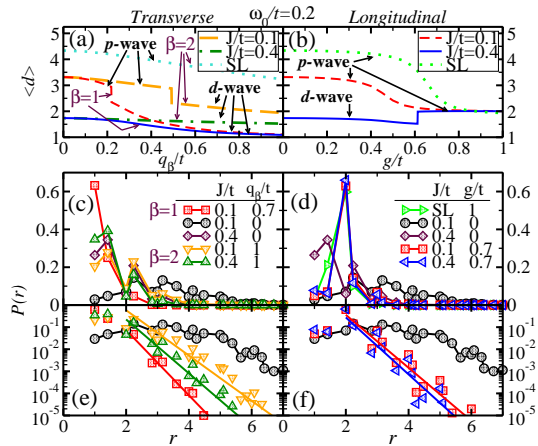}
\caption{(Color online) %
Average hole distance  $\langle  d \rangle $ computed at
$\omega_0/t=0.2$  and two different values of $J/t$ vs. { $q_{\beta}/t$;$\beta=1,2$   in (a) and vs. $g/t$ in
(b)}. Results for the system of spinless particles (SL) are shown with dotted
lines in (a) and (b); $P(r)$ at chosen values of $q_{1,2}/t$ in (c) and $g/t$ in
(d). $P(r)$ is normalized to $\sum_r P(r) = 1$; corresponding exponential
scalings of $P(r)$ for $r\gtrsim 2$ are shown  in (e) and (f).
}\label{fig3}
\end{figure}

Turning to linear EP coupling to  LP phonons we start  from $J/t = 0.1$. A $p-$state of two separate 
holes changes to a  bound bipolaron state  with increasing of $g/t$ at about
$g/t \sim 0.5$, see Fig.~\ref{fig3}(b). Transition to a bound state is not
sharp as in Fig.~\ref{fig3}(a) since there is no change of a symmetry.
Nevertheless, in the $N_h \to \infty$  limit, we anticipate  a sharp
transition from an unbound to a bound bipolaron state. Stabilization
of a $p-$wave state under the influence of LP
   is even more evident when starting from a $d-$wave
bound bipolaron state at $J/t = 0.4$. With increasing $g/t$, a change of symmetry  occurs around $g/t \sim 0.61$, from a
$d-$ to a $p-$wave state, see Figs.~\ref{fig3}(b,d,f).
A detailed 
inspection of a bound $p-$wave state in the regime  $g/t \gtrsim 0.61$
reveals unusually simple structure where the probability of
finding holes at a distance $r = 2$ is more than 0.6. This is
in a sharp contrast with the structure of a $d-$wave bound state
where $P(r = 2) < 0.1$ and the maximal value of $P(r)$ is at
$r = \sqrt{2}$, compare also Figs.~\ref{fig4}(a) and (b). 
{ Our calculations of hopping term modulated   by LP phonons,  as suggested in Ref.~\cite{ishihara}, as well leads to stabilization of a  bipolaron with a $p$-wave symmetry,  nonetheless, with  a distinct spacial structure. }

   { From  Fig.~\ref{fig3} is as well evident that linear coupling to TP leads a  stronger attraction between holes than coupling to  LP  as seen from Figs.~\ref{fig3}(e) and (f) that show steeper decay of $P(r)$ for TP at comparable values $q_1/t=g/t=0.7$.  At small $J/t=0.1$ a bound bipolaron state is obtained at unexpectedly small value of the dimensionless EP coupling constant $\lambda_q=q_1^2/4\omega_0t\sim 0.06$ in the TP case in contrast to   $\lambda_g=g^2/4\omega_0t\sim 0.31$ in the LP case.  } This result   suggests  that even a small pre-buckling within the CuO plane, that generates  non-zero linear EP coupling term to TP,  may have  a pronounced effect on the attraction between magnetic polarons.


\begin{figure}[htb]
\includegraphics[width=7cm]{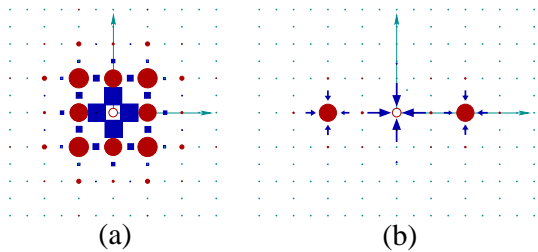}
\caption{(Color online) (a) ${\cal C}({\bf r})$ and ${\cal N}(\bf r)$ at $J/t=0.4$,  $q_2/t=1.0$, and $\beta=2$ ($d-$wave). Radii of circles, representing ${\cal C}({\bf r})$, located on Cu-sites are, proportional to the  probability of finding a hole-pair at a distance of $\bf r$. Empty circle is located at ${\bf r}=0$.  Sides  of squares, representing ${\cal N}(\bf r)$, located on O-sites,  are  proportional to  average numbers of phonon quanta at a distance $\bf r$ from a hole; (b) ${\cal C}({\bf r})$ and ${\cal X}(\bf r)$ at $J/t=0.4$ and  $g/t=0.7$ ($p-$wave). Lengths  of arrows, representing ${\cal X}(\bf r)$,  are proportional to displacements of O atoms along Cu-Cu bonds relative to the hole position. 
}\label{fig4}
\end{figure}

To investigate in more detail the nature of the magnetic-lattice bipolaron, we simultaneously present two correlation functions: hole-hole density 
$
{\cal C}({\bf r}) = \langle \sum_{ \bf i} n^h_{\bf i}n^h_{\bf i+r}
 \rangle,
$
and hole-phonon number 
$
{\cal N}({\bf r}) = \langle \sum_{ \bf i} n^h_{\bf i}a^\dagger_{\bf i+r}a_{\bf i+r}
 \rangle,
$ 
in Fig.~\ref{fig4}(a), for the case of a bound $d-$wave bipolaron state. Largest phonon numbers are found  at  the closest possible distance from the hole.
 The structure of ${\cal C}({\bf r})$ is consistent with $d_{x^2-y^2}$ symmetry despite its largest value at a distance of $r=\sqrt 2$, as already pointed out in Refs.~\cite{eder,riera2}.  In Fig.~\ref{fig4}(b) we show ${\cal C}({\bf r})$ and 
$
{\cal X}({\bf r}) = \langle \sum_{ \bf i} n^h_{\bf i}\left (a^\dagger_{\bf i+r}+a_{\bf i+r}\right )
 \rangle,
$ 
measuring displacements along Cu-Cu bonds relative to the position of the hole,  for the case of a $p-$wave ground state. Both correlations display the unidirectional spacial distribution.  
Correlation functions, presented in Figs.~\ref{fig4}(a) and (b), show detectable values only up to $r\lesssim 3$, despite maximal distance $l_{\mathrm max}=N_h+1=9$,  allowed in our calculations. 


{\it In conclusion} significantly  different bipolaron states  are found when EP coupling to either   TP or LP is switched on.  Linear  as well as quadratic EP coupling
to TP  stabilizes a $d-$wave bipolaron state.  The magnetic
background is  essential  for the formation
of a $d-$wave bipolaron. 
The effective bipolaron mass remains small in the case of quadratic EP coupling despite
lattice driven binding of the bipolaron.  

In contrast, increasing linear EP coupling to LP phonons changes the symmetry of a bound bipolaron from a $d-$wave state at zero EP coupling to a  $p-$wave state  followed by a substantial  change of the density-density correlation function. Since this state also has a large and anisotropic effective mass  and unidirectional spacial distribution we may speculate, that in a system with finite doping linear EP coupling to LP of O vibration  would lead to formation of stripe states. This finding is consistent with inelastic neutron  experiments showing strong coupling to the bond-streching mode in and around  the vicinity of the stripe phase in copper oxide superconductors \cite{reznik}.

J.B. acknowledges 
financial support of the SRA under grant P1-0044. 
S.M. and T.T. acknowledge the financial support of
the Next Generation Super Computing Project of Nanoscience
Program, CREST, and Grant-in-Aid for Scientific Research  from MEXT. This work was also supported by JPSJ and MHEST under the Japan-Slovenia Research Cooperative Program.
\vspace*{-0.5cm}
\bibliography{manubi}

\begin{thebibliography}{30}
\bibitem[{foo()}]{footnote}
\bibinfo{journal}{$d$-wave symmetry of a bipolaron is a many-body
  effect. A bound state of two particles with attractive interaction is always
  nodeless ($s$-wave in 2D), see also Ref.~[8]}.
\end{thebibliography}

\end{document}